\documentclass[a4paper,12pt,nonacm]{acmart}
\usepackage{graphicx}
\usepackage{tikz}
\usepackage{geometry}
\usepackage{xcolor}
\usepackage[T1]{fontenc}
\usepackage{tgbonum}
\usepackage{lmodern}
\usepackage{tgtermes}
\usepackage{tgpagella}
\usepackage{tgschola}
\usepackage{mathptmx}
\usepackage{utopia}
\usepackage{palatino}
\usepackage{bookman}
\usepackage{charter}
\usepackage{tabularx}
\usepackage{longtable}
\usepackage{hyperref}
\usepackage{titlesec}
\usepackage{fontawesome}
\usepackage{xcolor}

\definecolor{offwhite}{RGB}{255,255,255}

\newcommand{\papertitle}{Unfiltered Conversations: A Dataset of 2024 U.S. Presidential Election Discourse on Truth Social}
\newcommand{\paperauthors}{Kashish Shah, Patrick Gerard, Luca Luceri, Emilio Ferrara}
\newcommand{\paperaffiliation}{University of Southern California} 

\titlespacing*{\subsection}
{0pt}{*0.5}{*0.25} 
\begin{document}

\begin{titlepage}
    \begin{tikzpicture}[remember picture, overlay]
        \node[anchor=north west, inner sep=0] at (current page.north west) {
            \includegraphics[width=\paperwidth, height=\paperheight, trim=275 375 275 375]{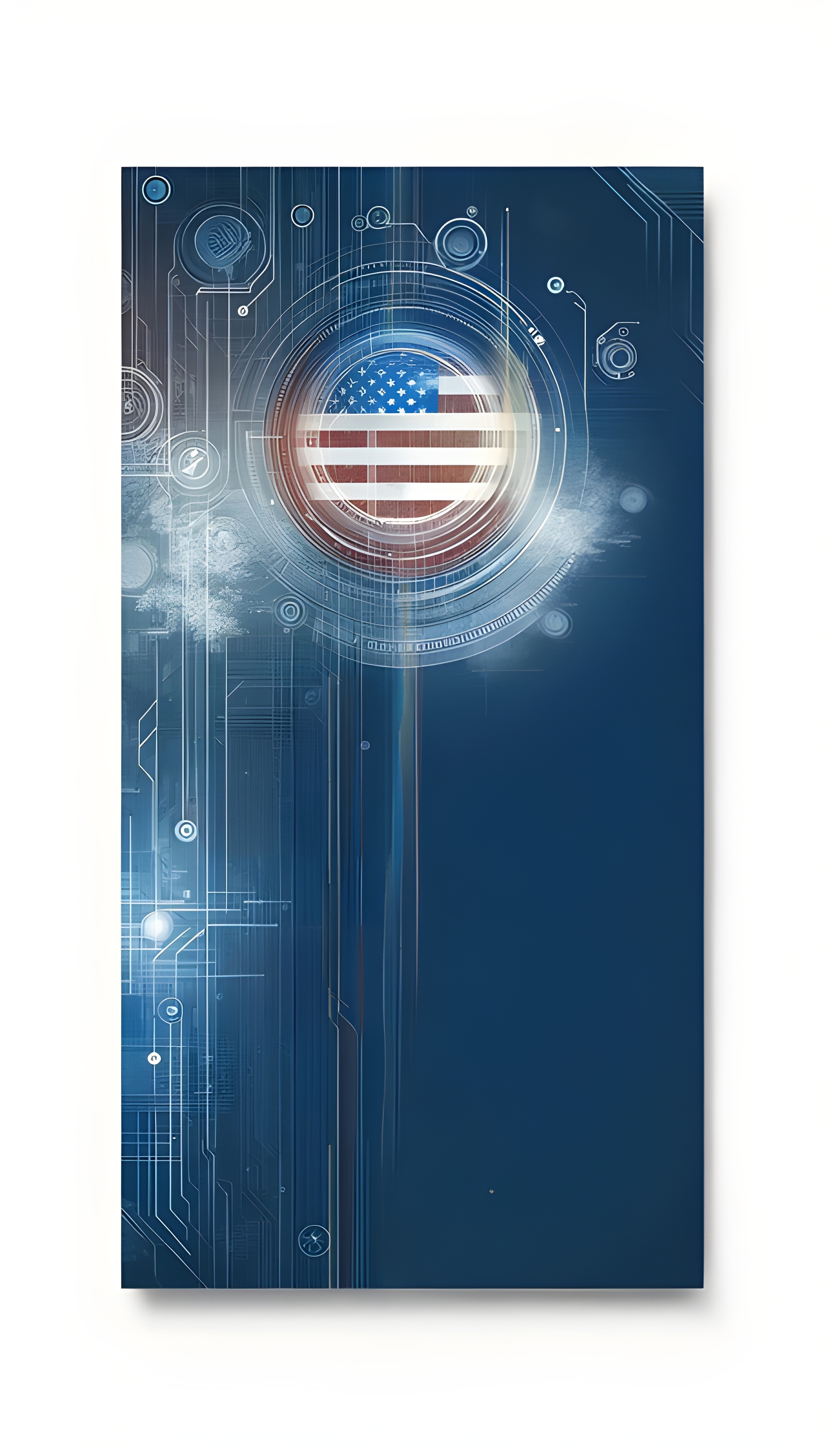}
        };
        \node[anchor=center, yshift=-9cm] at (current page.center) {
            \begin{minipage}{\textwidth}
                \raggedleft
                \color{offwhite}
                {\Huge \bfseries \fontfamily{qtm}\selectfont The 2024 Election Integrity Initiative }
                
                \vspace{1.5cm}
                
                {\LARGE \fontfamily{qtm}\selectfont \papertitle}
                
                \vspace{1.5cm}
                
                {\large \fontfamily{qtm}\selectfont \paperauthors}
                
                \vspace{1cm}
                
                {\Large \fontfamily{qtm}\selectfont \paperaffiliation}
                
                \vfill
                
                {\Large \fontfamily{qtm}\selectfont HUMANS Lab -- Working Paper No. 2024.8}
            \end{minipage}
        };
    \end{tikzpicture}
\end{titlepage}

\noindent{\LARGE \fontfamily{qtm}\selectfont \papertitle}

\vspace{0.5cm}

\noindent{\large \fontfamily{qtm}\selectfont \paperauthors}

\noindent{\large \fontfamily{qtm}\selectfont \textit{\paperaffiliation}}

\section*{Abstract}

Truth Social, launched as a social media platform with a focus on free speech, has become a prominent space for political discourse, attracting a user base with diverse, yet often conservative, viewpoints. As an emerging platform with minimal content moderation, Truth Social has facilitated discussions around contentious social and political issues but has also seen the spread of conspiratorial and hyper-partisan narratives. In this paper, we introduce and release a comprehensive dataset capturing activity on Truth Social related to the upcoming 2024 U.S. Presidential Election, including posts, replies, user interactions, content and media. This dataset comprises 1.5 million posts published between February, 2024 and October 2024, and encompasses key user engagement features and posts metadata. Data collection began in June 2024, though it includes posts published earlier, with the oldest post dating back to February 2022. This offers researchers a unique resource to study communication patterns, the formation of online communities, and the dissemination of information within Truth Social in the run-up to the election. By providing an in-depth view of Truth Social’s user dynamics and content distribution, this dataset aims to support further research on political discourse within an alt-tech social media platform. 
The dataset is publicly available at \url{https://github.com/kashish-s/TruthSocial_2024ElectionInitiative}

\section*{Introduction}

In recent years, the landscape of social media has seen the rise of “alt-tech”~\cite{dehghan2022politicization, vishnuprasad2024tracking} platforms that position themselves as alternatives to mainstream sites like Twitter/X and Facebook, often emphasizing free speech and minimal content moderation. Among these, Truth Social has become a dominant hub for right-wing social media users disillusioned with mainstream platforms’ moderation policies. Launched after former U.S. President Donald Trump's ban from major social media platforms due to his role in the January 6 U.S. Capitol attack,\footnote{\url{https://blog.twitter.com/en_us/topics/company/2020/suspension}} Truth Social combines the ethos of alt-tech platforms with the unique status of serving as a primary stage for a leading political figure. 

This combination of alt-tech structure and mainstream political figures has positioned Truth Social~\cite{gerard2023truth} as a powerful force within the social media ecosystem, establishing it as a prominent forum for political discourse, especially on contentious U.S.-related social and political issues. However, like other alt-tech ecosystems~\cite{cinelli2021echo}, Truth Social's lack of moderation---combined with a user-base self-selected along partisan lines---has helped facilitate the spread of conspiratorial, hyper-partisan, and fringe narratives~\cite{cinelli2022conspiracy, del2016spreading}. This has drawn increasing criticism,\footnote{\url{https://www.nytimes.com/interactive/2024/10/29/us/politics/trump-truth-social-conspiracy-theories.html}} raising important questions about the platform’s role in shaping public opinion and influencing democratic processes during this crucial electoral period.




Truth Social’s unique position in the social media ecosystem, particularly during the lead-up to the 2024 U.S. Presidential Election, makes it an important subject for study. As discussions intensify on platforms with minimal moderation, understanding the dynamics of user interaction and content distribution on Truth Social is essential for capturing how narratives and ideologies evolve around the election. To facilitate this research, we introduce a comprehensive dataset focused on the 2024 Election, capturing activity on Truth Social, posts, replies, user interactions, content and media. This dataset encompasses 1.5 million posts published between February 2022, and October 2024, with extensive metadata and user engagement metrics. Data collection began in June 2024, though it includes posts published earlier, with the oldest post dating back to February 2022. It provides researchers with a unique resource to study communication patterns, community formation, and the spread of information relevant to the election on Truth Social.

This dataset represents one of the most detailed public resources available for studying Truth Social’s role in the 2024 U.S. Election, offering insights into user dynamics, narrative propagation, and the formation of ideological communities within an alt-tech platform. With this dataset, researchers can examine Truth Social’s role in amplifying specific viewpoints, fostering communities aligned with political agendas, and potentially contributing to the polarization of online discourse surrounding the election. 
The aim of enabling scholars, policymakers, and media professionals to explore the complex interplay of free speech, minimal moderation, and political discourse on Truth Social during this pivotal election cycle inspired our data release.
The dataset is publicly available at \url{https://github.com/kashish-s/TruthSocial_2024ElectionInitiative}

\section*{Data Collection}
\begin{figure}[t]
  \centering
\includegraphics[width=0.8\textwidth]{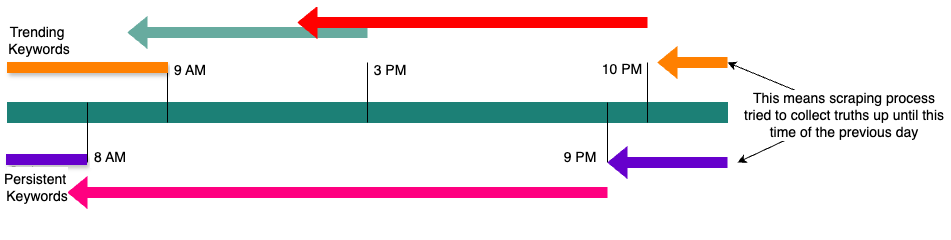}
  \caption{Data Collection Workflow}
  \label{fig:dataprocess}
\end{figure}

To capture trending narratives and persistent topics on Truth Social, we employ a daily data collection pipeline that gathers posts based on both dynamic and static keyword topics. These topics---referred to by Truth Social as both "topics" and "tags"---act as functional equivalents to hashtags. As illustrated in Figure \ref{fig:dataprocess}, the pipeline consists of two main components, which are detailed in the following. Each color in the figure represents a different scraping process and the time span that the process covers, i.e., the time window of the truths collected by that scraping process.

\subsection*{Trending Topic Collection}
Every day at 8am UTC and 9pm UTC, a Python script queries Truth Social’s API for the trending topics of the day. This is typically a list of 10-15 tags, each associated with a range of 500 to 20,000 posts. For each trending topic, the script gathers the most relevant posts. Truth Social's API tags posts with the topics they discuss and the script utilizes this same tagging to only collect posts that are central to each topic, ensuring that the data captures only the most important and popular discussions. For instance, if the trending tag is \#TrumpVance2024, the script accesses the endpoint \texttt{/tags/TrumpVance2024} to retrieve the posts. This approach allows us to simulate the experience of a user browsing trending topics on Truth Social, as it aligns with what Truth Social designates as significant within these topics.

We note that while a tag may be trending on day $t$, the posts returned may have been published on previous days ($t-1$, $t-2$, etc.). This asynchrony is built into the Truth Social API and provides a more comprehensive view of the historical activity contributing to current trends. Additionally, while tags are presented with casing, the API itself is case-insensitive, meaning capital letters do not affect query results.

\begin{table}[h!]
\centering
\caption{Keywords for Political Discourse Analysis}
\begin{tabular}{llll}
\hline
\addlinespace[1mm]
\textbf{Keywords} &  &  &  \\ \hline \addlinespace[1mm]
JoeBiden                & DonaldTrump            & 2024USElections       & USElections            \\
2024Elections           & 2024PresidentialElections & Biden               & JoeBiden               \\
JosephBiden             & Biden2024              & DonaldTrump           & Trump2024              \\
trumpsupporters         & trumptrain             & republicansoftiktok   & conservative           \\
MAGA                    & KAG                    & GOP                   & CPAC                   \\
NikkiHaley              & RonDeSantis            & RNC                   & democratsoftiktok      \\
thedemocrats            & DNC                    & KamalaHarris          & MarianneWilliamson     \\
DeanPhillips            & williamson2024         & phillips2024          & Democraticparty        \\
Republicanparty         & ThirdParty             & GreenParty            & IndependentParty       \\
NoLabels                & RFKJr                  & RobertF.KennedyJr.    & JillStein              \\
CornellWest             & ultramaga              & voteblue2024          & letsgobrandon          \\
bidenharris2024         & makeamericagreatagain  & VivekRamaswamy        &                        \\ 
\addlinespace[1mm]\hline
\end{tabular}
\label{table:keywords}
\end{table}

\subsection*{Persistent Topic Monitoring}
In addition to trending topics, we maintain a curated list of persistent keywords representing a broad range of political topics (detailed in Table~\ref{table:keywords}). This list serves as a consistent benchmark to track how communities on Truth Social engage with various political themes over time. At 9am UTC, 3pm UTC, and 10pm UTC each day, the script retrieves posts for each keyword in this list, accessing the corresponding `page' for each term, similar to the process used with trending topic. For each post associated with these keywords, we also collect the first-level comments (any subsequent comments to the first-level comments in the thread are not captured), capturing initial reactions and responses that provide further insight into the discourse surrounding these persistent topics. 

This dual approach---combining trending and persistent topic data---is designed to provide a longitudinal view of narrative dynamics on Truth Social. By tracking both emergent and established topics, we aim to gain insights into how conversations evolve and adapt over time; this may help reveal how topics resonate, transform, or fade within the platform’s user base in response to changing social and community contexts.

\section*{Data Processing}

After collecting raw data, we performed data cleaning to remove duplicates and irrelevant content, ensuring a high-quality dataset. We then aggregated keywords to highlight frequently discussed topics. The final dataset reflects the most prominent keywords in election conversations, providing a view of public engagement and trends on Truth Social leading up to the election. 
Notably, we will update the repository consistently for future collections as we collect more content. 


\section*{Exploratory Analysis}

\subsection*{Dataset Statistics}
In Table~\ref{tab:sumStats}, we present a summary of the statistics in the current version of the dataset presented in this paper.  Appriximately 40 thousand users were responsible for over 1.5 million posts, that accrued over 7.5 million likes and over 2.5 million \textit{retruths} (i.e., rebroadcasts).

\begin{table}[t]
\caption{Summary statistics of the dataset.}
\begin{tabular}{|l|l|}
    \hline
    Number  of truths& 561,450\\ \hline  
    Number  of  replies& 818,461\\ \hline
    Number  of likes   & 7,549,568\\ \hline
    Number  of retruths& 2,540,771\\ \hline
    Number  of unique users & 39,446\\ \hline
 Total number of entries&1,585,187\\\hline
\end{tabular}
\label{tab:sumStats}
\end{table}

\subsection*{Number of \textit{Truths} over time}
\begin{figure}[t]
\centering
\includegraphics[clip, trim=5 3 0 20, height=0.3\textheight]{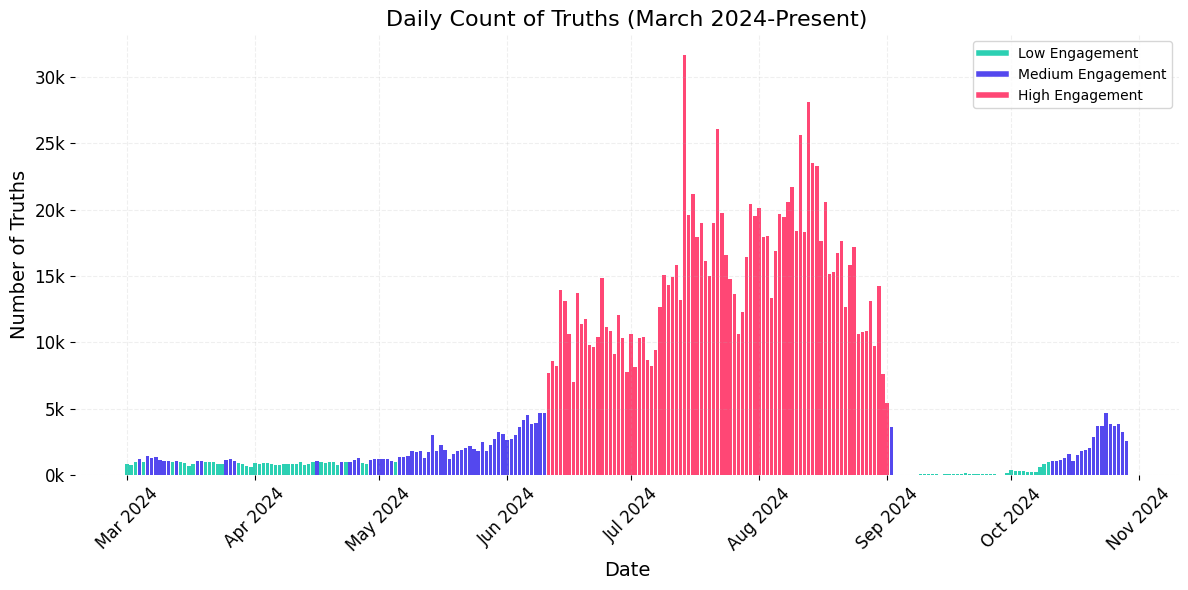}

\caption{Timeline of posts shared on Truth Social.}
\label{fig:videostime}
\end{figure}

In this first release, we collected posts between April 1, and September 1, 2024. Figure~\ref{fig:videostime} shows the volume of posts shared on Truth Social (also known as \textit{truths}) over time, with engagement levels assigned based on daily post counts, using quantiles to define low (0-33\%), medium (33-66\%), and high engagement(67-100\%). We note a rise in user activity as the election approaches. We note a rise in user activity as the election approaches.

\setlength{\belowcaptionskip}{-0.7cm}


\subsection*{Top Keywords}
\begin{figure}[b]
  \centering
\includegraphics[width=0.75\linewidth]{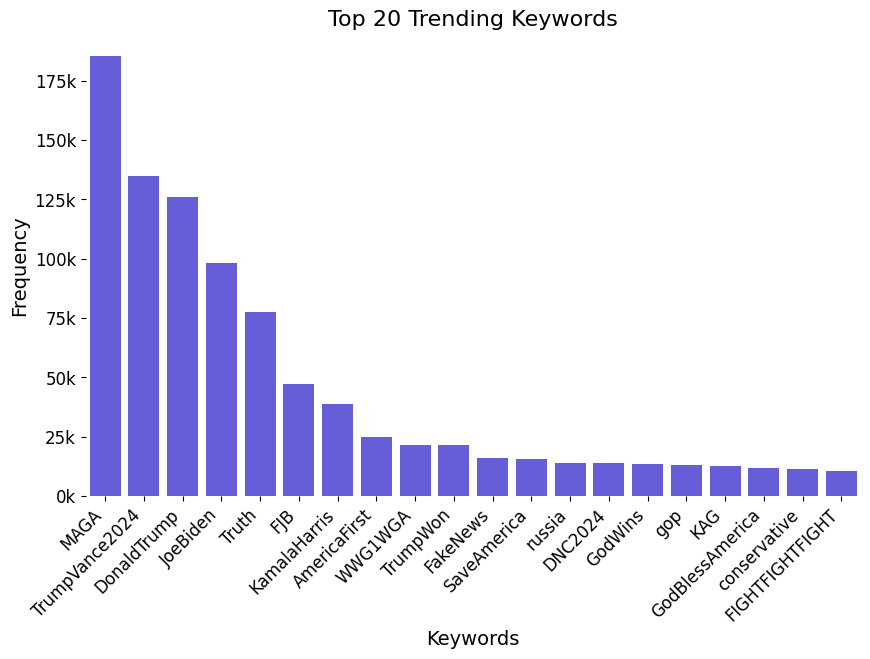}
  \caption{Top Keywords}
  \label{fig:top_phrases}
\end{figure}

 Figure~\ref{fig:top_phrases} displays the most recurring keywords and their frequencies in our dataset. The high frequency of certain terms indicates that the content primarily focuses on the presidential candidate Trump and his slogan (MAGA––Make America Great Again). Although Biden and Harris are frequently mentioned, the keyword "FJB"—an insulting acronym directed at President Biden—suggests a negative tone in messaging toward the Democratic candidates (Biden first, followed by Harris).

\begin{figure}[b]
    \centering
    \includegraphics[width=0.75\linewidth]{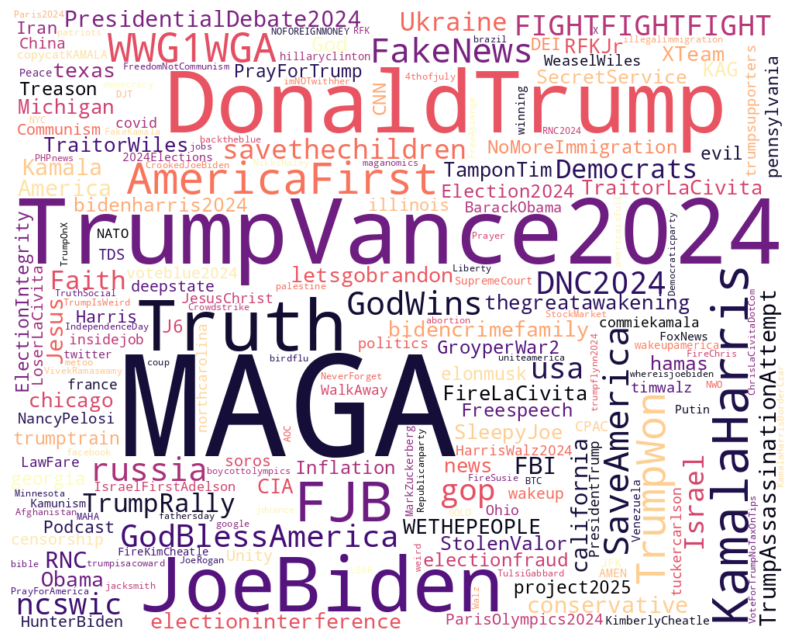}
    \caption{Prominent Election Keywords}
    \label{fig:prominent-keywords}
\end{figure}

Figure \ref{fig:prominent-keywords} visualizes all election-related keywords present in the dataset, with each term's size reflecting its popularity. Prominent terms like "MAGA", "TrumpVance2024", and "DonaldTrump" underscore a strong focus on Trump and his campaign. This diverse array of keywords not only highlights key political figures and slogans but also captures the polarized nature of election discussions on Truth Social, with significant engagement on both ideological and personal fronts. 

\begin{figure}[h]
    \centering
    \includegraphics[width=0.75\linewidth]{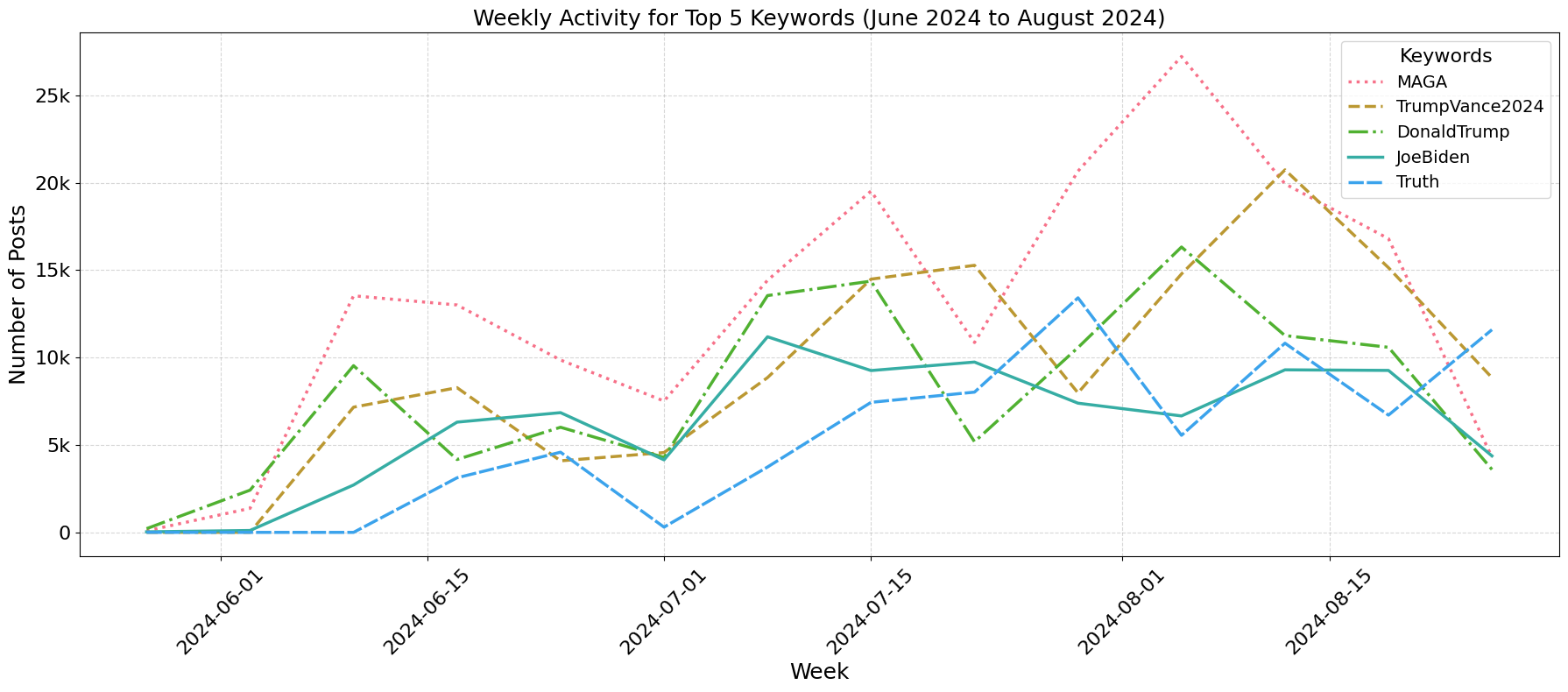}
    \caption{Keyword Engagement by Month}
    \label{fig:monthly-engagement}
\end{figure}

Figure \ref{fig:monthly-engagement} illustrates the weekly distribution of the most frequently used election-related keywords from June to August 2024. This time series plot shows a prominent focus on topics related to Trump's campaign, with keywords such as "MAGA," "TrumpVance2024," and "DonaldTrump" displaying significant peaks in engagement during June and August, possibly corresponding to key events or developments in the campaign. In contrast, mentions of Democratic figures, represented by keywords like "JoeBiden," maintain lower and more stable engagement levels throughout the period. This visualization provides insight into the dynamic shifts in public interest and sentiment on Truth Social, with engagement surges potentially linked to impactful campaign events or news cycles.

\section*{Conclusions}
This paper presents our current process for collecting Truth Social content related to the upcoming U.S. Presidential Election. Based on our preliminary exploratory study of 1.5 million posts shared on Truth Social, the most frequent keywords are associated with Donald Trump and the MAGA slogan. President Joe Biden is also frequently mentioned, occasionally alongside offensive tags.

This data collection enables a deeper understanding of political discourse on Truth Social and will continue after the election, tracking events as they unfold until the inauguration of the new president. 




\section*{Data Availability} 

Our data collection will continue uninterrupted for the foreseeable future. As the election approaches, we anticipate that the amount of data will grow significantly. 

A publicly accessible GitHub repository that we will continue to routinely update is available at \url{https://github.com/kashish-s/TruthSocial_2024ElectionInitiative}

\section*{About the Team}
The 2024 Election Integrity Initiative is led by Emilio Ferrara and Luca Luceri and carried out by a collective of USC students and volunteers whose contributions are instrumental to enable these studies. The authors are indebted to Srilatha Dama and Zhengan Pao for their help in bootstrapping this data collection. The authors are also grateful to the following HUMANS Lab's members for their tireless efforts on this project: Ashwin Balasubramanian, Leonardo Blas, Charles 'Duke' Bickham, Keith Burghardt, Sneha Chawan, Vishal Reddy Chintham, Eun Cheol Choi, Priyanka Dey, Isabel Epistelomogi, Saborni Kundu, Grace Li, Richard Peng, Gabriela Pinto, Jinhu Qi, Ameen Qureshi, Tanishq Salkar, Reuben Varghese, Siyi Zhou.
\textbf{Previous memos}: \cite{memo1, memo2, memo3, memo4, memo5, memo6, memo7}

\bibliography{memos,chatgpt,tiktok,genai}

\bibliographystyle{ACM-Reference-Format}
\newpage
\appendix


\end{document}